\documentclass[11pt]{amsart}
\usepackage{geometry}                
\usepackage{graphicx}
\usepackage{amssymb}
\usepackage{epstopdf}
\renewcommand{\div}{\mbox{div}}
\newcommand{\curl}{\mbox{curl}}
\newcommand{\R}{\mathbb{R}}
\newcommand{\Z}{\mathbb{Z}}
\newcommand{\T}{\mathbb{T}}
\renewcommand{\S}{\mathbb{S}}
\newcommand{\eps}{\varepsilon}
\newcommand{\pl}{\parallel}
\newcommand{\sgn}{\mbox{sgn}}

\title{Differential forms for plasma physics}
\author{R.S.MacKay}
\address{Mathematics Institute, University of Warwick, Coventry CV4 7AL, U.K.}
\email{R.S.MacKay@warwick.ac.uk}
\date{\today}                                           

\begin{document}

\begin{abstract}
Differential forms provide a coordinate-free way to express many quantities and relations in mathematical physics.  In particular, they are useful in plasma physics.  This tutorial gives a guide so that you can read the plasma physics literature that uses them and apply them yourself.
\end{abstract}

\maketitle

\section{Introduction}
Differential forms allow one to go beyond what vector calculus can cope with.  For example, they permit extension of the operator $B\cdot\nabla$ in a principled way to more general objects than scalar functions,
powerful generalisations of Gauss' and Stokes' theorems, and generalisations of 3D results like that a curl-free vector field is locally a gradient and a divergence-free one is locally a curl.  These extensions and generalisations are particularly useful in contexts like Hamiltonian dynamics, where the state space is usually not three-dimensional.  

One can write everything about differential forms in index notation (as in \cite{MTW}) and index notation generalises to arbitrary tensors (differential forms are only the antisymmetric covariant tensors).  But the concepts for differential forms are clearer without the reference to coordinate systems that index notation implies.  In particular one can see which relations involve an assumed volume-form or an assumed Riemannian metric and which are independent of these.

Differential forms were developed into a complete theory by Cartan in 1899.  Although they have been used to good effect in electromagnetics (see \cite{WR} for a recent review, including many pointers to the literature, of which I find \cite{De} particularly good) and in plasma physics (e.g.~\cite{CL}, who gave a tutorial appendix), many plasma physicists are unfamiliar with them.
The aim of this tutorial is to make them accessible to plasma physicists.

We give a definition of differential forms on a general oriented smooth manifold $M$.  Loosely speaking, a manifold is a topological space such that every point has a neighbourhood homeomorphic to either $\R^n$ or $\R_+ \times \R^{n-1}$ (the latter corresponding to boundary points of $M$).  The integer $n$ is called the dimension of the manifold.  A smooth manifold is a manifold for which the concepts of tangent vectors and differentiable functions $f:M\to \R$ are defined and satisfy desirable properties.  A more precise definition is given in an Appendix.  It is oriented if there is a continuous choice of set of $n$ linearly independent tangent vectors at each point (called a frame).
Examples of oriented smooth manifolds include $\R^n$, the circle, which I will consider as $\T = \R/\Z$, the $n$-torus $\T^n$ (which is the product of $n$ copies of the circle), and the $n$-sphere $\S^n$, which is the set of points in $\R_{n+1}$ such that $\sum_{i=0}^n x_i^2 = 1$ (note that, apart from $n=1$, it is not a product of $n$ copies of the circle).

For integer $k$ between $0$ and the dimension $n$ of the manifold (including $0$ and $n$), a {\em differential $k$-form} on $M$ is an antisymmetric $k$-linear map from ordered $k$-tuples of tangent vectors at any point of $M$ to the reals.  {\em Antisymmetric} means it changes sign under interchange of any pair of its arguments.  The case of a $0$-form is just a scalar function from $M$ to $\R$.  The case $k=n$ is called a {\em top-form}.
If we want to indicate a differential form $\omega$ at a point $p \in M$, we write $\omega_p$.
I shall assume whatever degree of differentiability is required for the results I shall state.

A good introduction to differential forms in the context of Hamiltonian mechanics is Ch.7 of \cite{Arn}.
A lot more advanced material of relevance to hydrodynamics and magnetohydrodynamics is given in \cite{AK}.

The paper starts by developing the main concepts for differential forms in the context of magnetic fields.  It then goes on to use them in charged particle dynamics.  It closes with a short discussion and an appendix on some additional topics.

\section{Magnetic fields}

A {\em magnetic field} $B$ is a volume-preserving 3D vector field.  

There are two ways to view a vector field on a manifold.  
The first is as a field of tangent vectors.  First define a differentiable parametrised curve to be a map $\gamma$ from $[-1,+1] \to M$ such that in one or any local coordinate system $\phi: U \to \R^n$ with $U$ a neighbourhood of the image of $\gamma$, $\phi \circ \gamma$ is a differentiable map from $[-1,+1] \to \R^n$.  Then a tangent vector is an equivalence class of differentiable parametrised curves through a point, where two such curves $\gamma_1, \gamma_2$ are considered equivalent if $d(\gamma_1(t),\gamma_2(t)) = o(t)$ as $t \to 0$, using any metric $d$ on $M$.  Intuitively, this says that $\gamma_1$ and $\gamma_2$ pass through the same point at $t=0$ and have the same velocity at $t=0$.  So a tangent vector is the velocity of a parametrised curve.  A vector field is a continuous field of tangent vectors; think of it as the velocity field of a fluid flow.

The second is as a first-order differential operator (``derivation") on scalar functions on the manifold, i.e.~a linear operator $L$ on the set of smooth functions $f:M \to \R$ that satisfies Leibniz' product rule $L(fg) = (Lf)g + f (Lg)$.  

The link is that a vector field $v$ defined as the velocity of a flow induces a first-order operator on differentiable functions $f: M\to \R$ by $v(f) = Df(v)$, where the derivative $Df_p$ at a point $p \in M$ is the linear map on tangent vectors $v$ at $p$ such that $f(p+\eps v) - f(p) = \eps Df_p(v) + o(\eps)$ as $\eps \to 0$ in any local coordinate system.
It is common to write $Df(v)$ as $v\cdot\nabla f$, but that notation suggests it depends on the choice of a Riemannian metric $g$ on $M$, because for vectors $a,b$, $a\cdot b$ represents $g(a,b)$ and $\nabla$ is defined using a Riemannian metric.  In fact, $\nabla$ is defined by the relation $v\cdot\nabla f = Df(v)$ for all vectors $v$, so the dependence on the metric cancels out. Thus I prefer to avoid the $v\cdot\nabla f$ notation.  

Every first order differential operator $L$ on functions is of the form $L(f) = Df(v)$ for some vector field $v$ in the first sense.  This can be proved using linearity and Leibniz' rule.

\subsection{Volume-preservation and magnetic flux-form}

The vector calculus way to write the volume-preservation condition is $\div\ B = 0$.  To explain how to write it in differential forms language will take several steps.

Firstly, we let $\Omega$ be the relevant volume-form in 3D.  A {\em volume-form} on a manifold is a non-degenerate top-form.  For a top-form, {\em non-degenerate} means that there exists an $n$-tuple on which it is non-zero.  In the 3D context, the interpretation is that $\Omega(\xi,\eta,\zeta)$ is the signed volume of the parallelopiped spanned by any ordered triple of tangent vectors $(\xi,\eta,\zeta)$.  In vector calculus language the standard Euclidean volume is $\Omega(\xi,\eta,\zeta) = \xi.(\eta \times \zeta)$.

The next step is to introduce the {\em flux-form} $\beta = i_B\Omega$ associated to $B$.  This is a 2-form giving the flux of $B$ through any infinitesimal parallelogram.  Specifically,
\begin{equation}
i_B\Omega (\xi,\eta) = \Omega(B,\xi,\eta).
\end{equation}
For a general vector field $X$, the {\em contraction operator} $i_X$ on an arbitrary differential $k$-form $\omega$ ($k>0$) is the $(k-1)$-form defined by putting $X$ as the first argument of $\omega$.  We extend to the case of $0$-forms $f$ by defining $i_X f=0$.
For future reference, {\em non-degeneracy} of a general $k$-form $\omega$ ($k>0$) is defined by $i_X \omega = 0 \implies X=0$.

For some purposes, in particular to work out the magnetic flux through a surface, $\beta$ is the more natural view of a magnetic field than $B$.
Given a 2-form $\beta$ and a volume-form $\Omega$ in 3D there is always an associated vector field $B$ such that $i_B \Omega = \beta$ and it is unique.  This is because $\Omega$ is assumed to be non-degenerate.

A $k$-form $\omega$ can be integrated over oriented submanifolds of dimension $k$.  Heuristically, we subdivide the submanifold into small parallelopipeds of dimension $k$ spanned by $k$ tangent vectors $\xi_1,\ldots \xi_k$, sum up $\omega(\xi_1,\ldots \xi_k)$ and take the limit as their size goes to zero.  For example $\int_S \beta$ gives the magnetic flux through a surface $S$ and $\int_V \Omega$ gives the volume of a region $V$.  Note that in this notation, no variable of integration is specified; this is because the formulation is coordinate-free.

For any $k$-form $\omega$, $0\le k < n$, there is a $(k+1)$-form $d\omega$ such that for any oriented $(k+1)$-submanifold $V$ with boundary $\partial V$ (with associated orientation),
\begin{equation}
\int_V d\omega = \int_{\partial V} \omega.
\end{equation}
This was called Stokes' theorem by Cartan, because it encompasses the usual Stokes' theorem (we will get to the differential forms version of $\curl$ later), but it also includes Gauss' theorem and Green's theorem.  The operator $d$, called {\em exterior derivative}, can be defined by
\begin{equation}
d\omega(\xi_1,\ldots \xi_{k+1}) = \lim_{\eps \to 0} \eps^{-k} \int_{\partial W_\eps} \omega,
\end{equation}
where $W_\eps$ is the parallelopiped spanned by $(\eps \xi_1,\ldots \eps \xi_{k+1})$ in some local coordinate system.  This definition of $d$ makes Stokes' theorem obvious.
On $0$-forms $f$ (scalar functions), $d$ is just the usual derivative:~applied to a vector $\xi$ it gives $df(\xi) = Df(\xi)$, the directional derivative along $\xi$.  As already discussed, this is commonly written as $\xi.\nabla f$.
The point of the definition of $d$ is to extend it to $k$-forms with $k>0$.
The definition of $d$ can be extended to act on top-forms by sending them all to 0.

Now consider $d\beta$ for $\beta = i_B\Omega$.  It is a 3-form in 3D.  The space of top-forms at a point is one-dimensional, and $\Omega$ is non-degenerate, so $d\beta$ is a multiple of $\Omega$.  $\div\ B$ is defined to be that multiple:
\begin{equation}
d\beta = \div B\ \Omega.
\end{equation}

Thus, in differential forms, the volume-preserving condition is written as $d\beta = 0$.  Any differential form whose exterior derivative is zero is called {\em closed}.  Thus the flux-form for a magnetic field is closed.

Note how the Cartan version of Stokes' theorem includes Gauss' theorem: for any vector field $X$, volume-form $\Omega$ and top-dimensional volume $V$, apply it to $i_X \Omega$ and use the definition of $\div$; it says $\int_V \div X\ \Omega = \int_{\partial V} i_X \Omega$.

\subsection{Magnetic 1-form}
In addition to the magnetic field $B$ and its flux-form $\beta = i_B \Omega$, it is useful to consider the associated 1-form $B^\flat$.  This is defined by
\begin{equation}
B^\flat (\xi) = B\cdot \xi
\end{equation}
on any tangent vector $\xi$.  Here $\cdot$ denotes the inner product of two vectors.  For a general Riemannian metric $g$, this is defined by $a\cdot b = g(a,b)$.
Equivalently we can write $B^\flat = i_B g$, by extending the contraction notation from differential forms to any covariant tensor, such as a metric tensor (a {\em covariant tensor} is a multilinear map from the tangent space to the reals, without the antisymmetry condition).

The inverse operation to $^\flat$ is denoted $^\sharp$, taking a 1-form to a vector field.  It is well-defined because any Riemannian metric is assumed to be non-degenerate (in fact positive definite, but one can extend both operations to Lorentzian metrics too).

It is also convenient to define
$|B| = \sqrt{g(B,B)}$.
Using the above, one can write $|B|^2 = i_B B^\flat$.

Finally we revisit the relation $df(v) = v\cdot\nabla f$ for a scalar function $f$.  Using the above notation we see we can write
\begin{equation}
(\nabla f)^\flat= df, \quad \nabla f = df^\sharp.
\end{equation}

\section{Cross product}
The cross product of vectors enters many formulae involving the magnetic field, e.g.~the Lorentz force\footnote{Some call it Laplace force.} $F = J \times B$, where $J$ is the current-density vector field.  The cross product is specific to 3D.  There are two ways of writing it in differential forms, which are equivalent iff the assumed volume-form $\Omega$ is {\em natural} for the Riemannian metric $g$, i.e.~applied to any orthonormal ordered triple, $\Omega$ produces $\pm 1$ according to the orientation of the triple.

The first way is that $J \times B$ is the vector such that 
\begin{equation}
(J\times B)^\flat = i_B i_J \Omega.
\end{equation}
Note the reversal of order and the explicit dependence of the cross product on a volume-form $\Omega$ and a Riemannian metric (through $^\flat$).
One could write $J\times B=(i_B i_J \Omega)^\sharp$, but it is natural to consider forces as covectors rather than vectors, because the work done by a force vector $F$ moving through a displacement $\xi$ is $F^\flat(\xi)$.  So I prefer the first formulation.  Similarly, momenta are naturally covectors $p$, with Newton's law $\dot{p}=F^\flat$ and $p(\dot{q})$ tells how fast the action integral changes for velocity $\dot{q}$ of configuration.

The second way is that
\begin{equation}
i_{J\times B} \Omega = J^\flat \wedge B^\flat,
\end{equation}
where the {\em wedge product} of a $k$-form $\omega$ and an $m$-form $\eta$ is defined to be the $(k+m)$-form
\begin{equation}
(\omega \wedge \eta)(\xi_1,\ldots \xi_{k+m}) = \sum_{\sigma \in Sh(k,m)} \sgn(\sigma)\ \omega(\xi_{\sigma(1)},\ldots \xi_{\sigma(k)}) \ \eta(\xi_{\sigma(k+1)},\ldots \xi_{\sigma(k+m)}),
\end{equation}
with $Sh(k,m)$ being the set of permutations $\sigma$ of $\{1,\ldots k+m\}$ such that $\sigma(1)<\ldots \sigma(k)$ and $\sigma(k+1)<\ldots \sigma(k+m)$ ({\em shuffles}).
This definition of $J\times B$ again depends explicitly on a volume-form and a Riemannian metric.

It is worth mentioning at this stage how $d$ and $i_B$ act on wedge products:
\begin{equation}
d(\omega \wedge \eta) = d\omega \wedge \eta + (-)^k \omega \wedge d\eta,
\end{equation}
where $k$ is the degree of $\omega$, and the same for $i_B$.
Note the case of the wedge product with a 0-form is just multiplication: $f \wedge \eta = f \eta = \eta \wedge f$.

\section{curl and de Rham cohomology}
\label{sec:curl}

Written in differential forms, the curl of a 3D vector field $B$ is the vector field $J$ such that
\begin{equation}
i_J \Omega = dB^\flat.
\end{equation}
So it depends on both the volume-form $\Omega$ and the Riemannian metric (via $^\flat$).  We see that the flux-form $j = i_J\Omega$ is the natural object here.  

Note that the above expresses the relation between a current density $J$ and a magnetic field $B$ in units for which the magnetic permeability $\mu_0=1$.  We shall take that convention throughout.

A fundamental ingredient of the theory of differential forms is that $d^2=0$.  One way to see this is if $\alpha$ is an arbitrary $(k-1)$-form, then let $\beta = d\alpha$.  Applying Stokes' theorem twice starting with an arbitrary $(k+1)$-submanifold $V$ gives $\int_V d\beta = \int_{\partial V} \beta = \int_{\partial\partial V} \alpha = 0$ because the boundary of a boundary is empty.  So $d^2\alpha = d\beta = 0$.

So for example, the flux-form $j$ defined above is automatically closed ($dj=0$).  This expresses that div of a curl is zero.
I used the notation $B$ for an arbitrary 3D vector field, which might suggest that I was assuming it is volume-preserving, but that was not used above.

As a second example, if $B$ is  a gradient vector field $\nabla f$ then its curl is automatically zero.  This is because $(\nabla f)^\flat = df$, so $j=d^2f = 0$.

Now the question arises for a differential $k$-form $\beta$, if $d\beta = 0$ then is $\beta = d\alpha$ for some $(k-1)$-form $\alpha$? In the positive case, $\beta$ is called {\em exact}.  In a contractible subset of a manifold this is true ({\em Poincar\'e's lemma}).  It generalises the standard results that (i) if $\curl\ v=0$ then $v$ is locally the gradient of some function $\phi$, and (ii) if $\div\ B = 0$ then $B$ is locally $\curl\ A$ for some vector field $A$.  The latter is usually derived as a global result in $\R^3$ via Helmholtz' theorem, but 
Poincar\'e's lemma can be proved in a bounded contractible subset, analogous to the standard way for proving (i) (see Appendix).


There can be global topological obstructions, however.
For example, for the flux-form $\beta = i_B\Omega$ for a volume-preserving field $B$ we have $\beta$ is closed.  So locally, there is a 1-form $a$ such that $\beta = da$.  $A= a^\sharp$ is known as a {\em vector potential} 
\footnote{It would be more natural to regard the 1-form $a$ as a potential for $B$, and this is done in some of the literature and is then often denoted by $A$, but I stick with plasma physics convention.} 
for $B$, so the relation is that $\beta = dA^\flat$.  But if there is a (non-contractible) closed surface $S$ over which $\int_S \beta \ne 0$ then there can not be a globally defined vector potential for $B$.  In practice for physical magnetic fields this appears never to occur, though maybe in a big enough piece of the universe we will find it does.  Note that the vector potential for $B$ is not unique: one can add any gradient to $A$.

As a second example, however, suppose $B$ is a {\em vacuum magnetic field}, i.e.~$\curl\ B = 0$, equivalently $dB^\flat=0$.  Then $B^\flat$ is locally the derivative of a $0$-form (i.e.~scalar function) $\phi$, i.e.~$B^\flat = d\phi$ or equivalently $B = \nabla \phi$.  But it may well fail to be a gradient globally.  For example, if the region in which $B$ is a vacuum field is a solid torus, as would arise by generation of $B$ by currents in coils surrounding it, then for any closed curve $\gamma$ going the non-contractible way in the solid torus, $\int_\gamma B^\flat = \int_{\mathcal{A}} j$ where $\mathcal{A}$ is a disc in the surrounding 3D space spanning $\gamma$, and this is the total external current $I_{ext}$ through the hole in the solid torus, whereas if $B^\flat = d\phi$ then its integral along any closed curve is zero.  In this example there is an easy solution: $B=\nabla \phi$ in the solid torus for a multivalued function $\phi$, which increases by $I_{ext}$ for each revolution around the solid torus.

The discussion above introduces the fascinating topic of de Rham cohomology.  The $k^{th}$ {\em de Rham cohomology group} $H^k$ of a manifold $M$ is defined to be the quotient of the set of all closed $k$-forms on $M$ by the set of all exact ones, i.e.~consider two closed forms to be equivalent if they differ by an exact one.  It is a group under addition.  Actually it is a real vector space.  Its dimension $\beta_k(M)$ is called the $k^{th}$ {\em Betti number}.  $H^1$ will play a role when we come to the Hamiltonian version of Noether's theorem.

Now that we have learnt about the wedge product and Poincar\'e's lemma we can formulate the Clebsch representation of a volume-preserving vector field in differential forms, namely $d\beta=0$, for $\beta$ the flux 2-form, implies $\beta = da$ locally, for some 1-form $a$, as above.  Every 1-form $a$ can be written locally as $f dg$ for some functions $f$ and $g$.  So $\beta = d(fdg) = df \wedge dg$, which is the differential forms version of $B=\nabla f \times \nabla g$ in 3D.  There are usually global topological obstructions, however.

\section{Lie derivative}
\label{sec:Lie}

The Lie derivative $L_B$ of differential forms (or more general tensors, to be defined later in this section) along a vector field $B$ is one of the most useful concepts, describing how the differential form changes as viewed along the flow of the vector field.
I call the vector field $B$, but no 3D or volume-preserving conditions are assumed.

We begin with the simple case of a $0$-form, i.e.~scalar function $f$.  Then $L_Bf$ is defined by
\begin{equation}
L_B f = df(B) = i_B df = Df(B) = B\cdot\nabla f.
\end{equation}
So $L_B$ is just the first-order operator view of the vector field $B$.  It is so important in plasma physics, however, that my PhD supervisor John Greene chose car registration B$\cdot$GRAD (using a screw for the dot). Equations of the form $L_B f = g$ with vector field $B$ and function $g$ given are called {\em magnetic differential equations}.

The real power of the Lie derivative is to describe how fast other objects like $k$-forms with $k>0$ or vector fields change along a vector field $B$.  In Euclidean space one can get away with writing expressions like $B\cdot\nabla B$ but they require careful interpretation (e.g.~in curvilinear coordinates) and do not always behave how you might expect.  The nice way to define how fast a general tensor field varies along a vector field is via the flow of the vector field.

Given a smooth vector field $B$ on $M$, it defines a {\em flow} $\phi: \R \times M  \to M$, $(t,x) \mapsto \phi_t(x)$ (at least locally around $t=0$) by
\begin{equation}
\partial_t \phi_t(x) = B(\phi_t(x)),
\end{equation}
with $\phi_0(x)=x$.  As a shorthand, write the derivative $D\phi_t$ with respect to $x\in M$ as $\phi_{t*}$ (note the asterisk is subscripted, indicating that it pushes tangent vectors forward).
Define the {\em pullback operator} $\phi_t^*$ of the differentiable map $\phi_t$ on differential forms (note superscript), e.g.~$k$-form $\omega$, by
\begin{equation}
(\phi^*_t \omega)_p(\xi_1,\ldots \xi_k) = \omega_{\phi_t(p)}(\phi_{t*}\xi_1,\ldots \phi_{t*}\xi_k).
\label{eq:pullback}
\end{equation}
Define
\begin{equation}
L_B \omega = \partial_t \phi^*_t\omega |_{t=0}.
\label{eq:Liecovar}
\end{equation}
Cartan proved the ``magic formula''
\begin{equation}
L_B = i_B d + d i_B
\end{equation}
for $L_B$ on differential forms.
We recover the above definition on $0$-forms, $L_B = i_B d$, because of the convention that $i_B =0$ on functions.
For top-forms, we see that $L_B = d i_B$ because of the convention that $d = 0$ on top-forms.  Thus an alternative way to write the definition of $\div$ with respect to a chosen volume-form $\Omega$ is 
\begin{equation}
L_B \Omega = \div B\ \Omega.
\end{equation}

One of the main points of the Lie derivative is to articulate how the integral of a form over a surface moving with the flow of a vector field $B$ evolves.  The answer is
\begin{equation}
\partial_t \int_{\phi_t S} \omega = \int_{\phi_t S} L_B \omega.
\end{equation}
The two terms of Cartan's magic formula capture how $\omega$ changes as viewed along the flow and the effect of how the boundary of $\phi_t S$ changes.

Some nice properties of the Lie derivative on forms are:
\begin{eqnarray}
d L_B &=& L_B d \\
L_B(\alpha \wedge \beta) &=& (L_B\alpha) \wedge \beta + \alpha \wedge L_B\beta \label{eq:Lwedge}\\
L_{\lambda B} \omega &=& \lambda L_B \omega + d\lambda \wedge i_B \omega. \label{eq:LlambdaB}
\end{eqnarray}
Note the case of (\ref{eq:Lwedge}) where $\alpha$ is a $0$-form $f$ gives 
\begin{equation}
L_u (f\beta) = (L_u f) \beta + f L_u\beta,
\end{equation}
and the case of (\ref{eq:LlambdaB}) where $\omega$ is a 0-form $f$ gives
\begin{equation}
L_{\lambda B} f = \lambda L_B f.
\end{equation}

The definition (\ref{eq:Liecovar}) extends to all {\em covariant tensors}, defined to be multilinear maps on ordered sets of vectors.  An example that is not a differential form is a metric tensor, because it is symmetric rather than antisymmetric.  The magic formula does not work, however, for general covariant tensors. 

It is convenient to extend the definition of the Lie derivative to {\em contravariant tensors} (multilinear maps on ordered sets of covectors) and {\em mixed tensors} (multilinear maps on ordered sets of vectors and covectors) too.   The main example that will concern us is just vector fields, so we restrict attention to that case.
A formula to compute the Lie derivative of a general tensor is given in the Appendix.
The pullback of a vector field $Y$ under the flow $\phi_t$ of vector field $B$ is
\begin{equation}
\phi^*_t Y|_p = \phi_{-t*} Y |_{\phi_t(p)}.
\end{equation}
Then we define
\begin{equation}
L_B Y = \partial_t \phi_t^*Y|_{t=0}.
\end{equation}

An alternative way to write $L_BY$ is the commutator $[B,Y]$.  To understand this, think of a vector field $X$ as the associated first order operator $i_Xd$ on functions.  Then $[B,Y] = BY - YB$ is interpreted as $i_Bdi_Yd - i_Ydi_B d$.  Although this looks second order, the second-derivative terms cancel and it is actually first order and equals $L_B Y$.
An alternative way to think of the commutator $[B,Y]$ is to flow from a point $p$ for time $t$ along $B$ followed by time $s$ along $Y$ and compare the result with flowing from $p$ along $Y$ for time $s$ followed by $B$ for time $t$.  The difference in any local coordinate system is of order $st$ and if one takes the limit of the quotient as $s,t \to 0$ one obtains a vector at $p$, which we call $[B,Y]$.
In vector calculus language, $[B,Y] = B\cdot\nabla Y - Y\cdot\nabla B$.
The commutator is antisymmetric.

A useful relation is that on differential forms
\begin{equation}
i_{[J,B]} = i_J L_B - L_B i_J = L_J i_B - i_B L_J,
\label{eq:icomm}
\end{equation}
which we shall use shortly.

We give some examples of how the Lie derivative shows up in plasma physics.  I say a {\em magnetohydrostatic (MHS) field} \footnote{This is not universal terminology.  The concept of MHS field lies between a magnetostatic field, being a magnetic field resulting from a prescribed steady current distribution ($\curl\ B = J, \div\ B = 0$), and a magnetohydrodynamic equilibrium, which allows steady mean flow of plasma and anisotropic pressure.} 
is a 3D volume-preserving field $B$ such that $J\times B = \nabla p$ for some scalar function $p$, where $J = \curl\ B$.
In differential forms these equations are written as $i_Bi_J\Omega = dp$ and $i_J\Omega = dB^\flat$.  One can eliminate mention of $J$ to obtain $i_BdB^\flat = dp$.  Now $di_B B^\flat = d|B|^2$, so it can also be written as 
\begin{equation}
L_B B^\flat = d(p+|B|^2).
\end{equation}
Continuing further, apply (\ref{eq:icomm}) to $\Omega$ to obtain
\begin{equation}
i_{[J,B]}\Omega = i_JL_B\Omega - L_B i_J\Omega.
\end{equation}
Now $L_B\Omega=0$ by $\div\ B=0$.  Also $i_J\Omega = dB^\flat$ and $d$ commutes with $L_B$ on forms, so
\begin{equation}
i_{[J,B]}\Omega = -dL_B B^\flat = -d^2(p+|B|^2) = 0
\end{equation}
for a MHS field.  As $\Omega$ is non-degenerate, we deduce that $[J,B]=0$.

Conversely, for a 3D volume-preserving vector field $B$ that commutes with its curl, $J$, there is a possibly multivalued scalar function $p$ such that $J\times B = \nabla p$.  This is because the hypotheses imply $dL_B B^\flat=0$ so $L_B B^\flat$ is locally $df$ for some function $f$ and then we can set $p = f - |B|^2$.  One can interpret solutions of $[J,B]=0$ as MHS fields subject to a possible additional force round each non-contractible loop, like an electromotive force for a charged plasma.

\section{So What?}

A nice consequence of the equation $i_Bi_J\Omega = dp$ for MHS fields and the resulting commutation relation $[J,B]=0$ is that bounded regular components of level sets of $p$ are tori, $B$ and $J$ are tangent to them, and the flows of $B$ and $J$ are simultaneously conjugate to rotations of different winding ratios on them.
A component of a level set of $p$ is {\em regular} if $dp \ne 0$ everywhere on it.
The flow $\phi^B$ of $B$ on a 2-torus $T$ is {\em conjugate to a rotation} if there exists $h: T \to \R^2/\Z^2$ such that $h(\phi^B_t(x)) = h(x) + r t$ for all $x \in T$, for some $r \in \R^2$, called the {\em rotation vector}.  The map $h$ is called a {\em conjugacy}.  Simultaneous conjugacy for $B$ and $J$ means they use the same conjugacy but may have different rotation vectors.
These results are known, but the following derivation is slick.

Firstly, a regular component of a level set of any scalar function in 3D is a 2D submanifold.  Secondly, applying $i_B$ or $i_J$ to $i_Bi_J\Omega = dp$ produces $i_B dp=0$ and $i_J dp = 0$ because $\Omega$ is antisymmetric, so $B$ and $J$ are tangent to the regular level sets of $p$ (this is the same as taking the inner product of $J\times B = \nabla p$ with $B$ or $J$, but gets rid of the irrelevant apparent dependency on a Riemannian metric).  Thirdly, if $dp \ne 0$ everywhere on a submanifold and $i_Bi_J\Omega = dp$ then $J$ and $B$ are independent everywhere on it.  
Finally, the Arnol'd-Liouville theorem (e.g.~\cite{Arn}) implies that for a bounded 2D submanifold invariant under the flows of two commuting vector fields that are independent everywhere on it, the submanifold is a torus and the flows are simultaneously conjugate to rotations of different winding ratios on them.  The latter conclusion is along the lines of the construction of magnetic coordinates such as Boozer and Hamada coordinates but makes it more natural.  Indeed,  $[J,B]=0$ leads directly to Hamada coordinates.

The Arnol'd-Liouville theorem is usually presented for integrable Hamiltonian systems, but the above key step applies more generally.  Note that many people deduce the submanifold is a torus just from its being bounded, supporting a nowhere-zero vector field (both $J$ and $B$ satisfy this as it is a consequence of their being independent), and orientable (which is a consequence of being a regular component of a level set of a function in an orientable space).  One just computes the Euler characteristic of the submanifold to be zero by Poincar\'e's index theorem and then uses the classification of surfaces.  That proof does not, however, give the additional information that $J$ and $B$ are simultaneously conjugate to rotations.

The winding ratio $\iota$ of the magnetic field on a flux surface can be expressed nicely using differential forms.  It is defined as the long-time limit of the ratio of the number of turns a fieldline makes in the poloidal direction to the number in the toroidal direction.  So it is the ratio of the components of the rotation vector.  In differential forms,
\begin{equation}
\iota = -\frac{\int_{\gamma_t} i_Bi_n\Omega}{\int_{\gamma_p} i_Bi_n\Omega},
\end{equation}
where $n = \nabla\psi/|\nabla\psi|^2$, over any cycles $\gamma_j$ on the flux surface such that $\gamma_t$ makes one toroidal turn and no poloidal ones and vice versa for $\gamma_p$.

\section{Applying diffeomorphisms}

If $B$ is a vector field on a manifold $M$, or $\omega$ a differential form on it, and $\varphi$ is a diffeomorphism from $M$ to another manifold $N$ (or possibly the same one), there is a natural way to produce a corresponding vector field and differential form on $N$.
Firstly, let $\varphi_*$ denote the derivative $D\varphi$ of $\varphi$, as in Section~\ref{sec:Lie}.  Then $\tilde{B} = \varphi_*B$ is a vector field on $N$.  As an example, $h$ a conjugacy of the flow of $B$ on a torus $T$ to a rotation says $h_*B = r$, constant on $\R^2/\Z^2$.
One might think that $\varphi_*$ could be used to make new magnetic fields from old ones, but the volume-preservation condition might not be satisfied.  If we suppose $B$ preserves volume-form $\Omega$ on $M$ then $di_B\Omega = 0$.  Applying $\varphi_*$ to this we obtain $di_{\tilde{B}}\varphi_*\Omega$, where the definition of $\varphi_*$ has been extended to the {\em push-forward} on $k$-forms $\omega$ by
\begin{equation}
(\varphi_*\omega)_p(\xi_1,\ldots \xi_k) = \Omega_{\varphi^{-1}(p)}(\varphi_*^{-1}\xi_1,\ldots \varphi_*^{-1}\xi_k),
\end{equation}
and we note that $\varphi_*$ passes through $d$ and through contractions (in the sense that $\varphi_*i_B\Omega = i_{\varphi_*B}\varphi_*\Omega$).
Thus $\tilde{B}$ preserves a volume-form on $N$, but perhaps not the one you had in mind.
If one has a prior choice $\tilde{\Omega}$ of volume-form on $N$ then $\tilde{B}$ preserves it iff $\varphi_*\Omega = \tilde{\Omega}$.

Let us consider instead the action of $\varphi_*$ on the magnetic flux-form $\beta = i_B\Omega$.  We know that $\div\ B = 0$ iff $d\beta=0$.  We let $\tilde{\beta} = \varphi_*\beta$ on $N$.  Applying $\varphi_*$ to $d\beta=0$ we deduce that $d\tilde{\beta}=0$.  So $\varphi_*$ takes a magnetic flux-form to one on $N$, without any further conditions.  This is one of the reasons to consider the magnetic flux-form to be more fundamental than the magnetic field.

Such diffeomorphisms $\varphi$ (with $N=M$) arise as the flow $\phi_t$ of the velocity field $v$ of a perfectly conducting fluid (which can be time-dependent).  The induction equation $\partial_t B =\curl(v\times B)$ implies $\partial_t \beta = -L_v \beta$ for the magnetic flux-form.  The definition of $L_v$ on forms used the pullback, but one can equally use the pushforward with change of sign.  So the flow applies $\phi_{t*}$ to $\beta$.  This is Alfven's {\em frozen flux theorem}.  

It is common to restrict attention to incompressible flows, in which case the induction equation implies that $\partial_t B = [v,B]$ so $\phi_{t*}$ then also gives the evolution of $B$.  This is often referred to as preservation of the topology of the magnetic field, but is much stronger than that term suggests, because of Alfven's theorem.  In general the induction equation for $B$ is $\partial_t B = -[v,B] - (\div\ v) B$ so there is an additional stretching effect on $B$ from convergence of $v$, beyond that implied by $\phi_{t*}$.

\section{Coordinates}

The point of differential forms is to be coordinate-free.  If you are desperate to connect to index notation, however, here is the correspondence.

A {\em coordinate system} or {\em chart} on an open subset $U$ of a manifold $M$ of dimension $n$ is a set of $n$ functions $x^1,\ldots x^n : U \to \R$ such that the map $p \mapsto (x^1(p),\ldots x^n(p))$ is a diffeomorphism from $U$ to its image in $\R^n$.  It is conventional to use superscripts for the coordinate functions.  

The linear operators $\partial_i = \frac{\partial}{\partial x^i}$ (keeping the other $x^j$ fixed) form a basis for the tangent space $T_pM$ at a point $p \in U$ (in the differential operator view of vectors).  It is conventional to use subscripts for these partial derivative operators.
So a vector $B$ at $p$ can be expanded as $B = \sum_{i=1}^n B^i \partial_i$.  It is conventional to use superscripts for the components of a vector.  It is convenient to adopt the {\em summation convention} that in an expression with an index appearing once as a subscript and once as a superscript, there is an implied sum over the possible values of that index, so $B = B^i\partial_i$, for example.

The 1-forms $dx^i$ at a point $p$ form a basis of $T^*_pM$, so $B^\flat$ can be expanded as $B^\flat = B_i dx^i$, using subscripts for its components, and summation convention.

It is common to think of a vector field $B$ and its associated 1-form $B^\flat$ as being the same object and refer to the $B^i$ as being its contravariant components and the $B_i$ as its covariant components.  They are related, as in the definition of $^\flat$, by an assumed Riemannian metric.  A metric tensor $g$ can be expanded as $g(\xi,\eta) = g_{ij}\xi^i\eta^j$ on pairs of vectors $\xi,\eta$ with respect to their components in a coordinate system.  Then the relation between $B$ and $B^\flat$ can be written $B_i = g_{ij} B^j$.
The notation $^\flat$ corresponds to the operation of lowering indices in index notation.  The operation $^\sharp$ corresponds to raising them.  If $\alpha$ is a 1-form $\alpha_i dx^i$ then $\alpha^\sharp$ is a vector field $\alpha^i \partial_i$ with $\alpha^i = g^{ij}\alpha_j$, the matrix $g^{ij}$ being the inverse of the matrix $g_{ij}$.

A minor caution: many plasma physicists use the term ``contravariant representation" of a vector field $B$ to mean the flux-form $\beta$, rather than the vector field itself.  They agree, however, that the ``covariant representation" of a vector field $B$ is the 1-form $B^\flat$.

A basis for 2-forms at a point is given by the $dx^i \wedge dx^j$ with $i<j$ to avoid duplication.  It can be tidier to think of the basis elements as being $dx^i \wedge dx^j - dx^j \wedge dx^i$ with $i<j$, which are twice the preceding ones.  So one can expand a 2-form like $\beta$ as $\beta = \beta_{ij}dx^i \wedge dx^j$, with $\beta_{ij}$ antisymmetric.  Similarly, a volume-form $\Omega$ in 3D can be written as $\Omega = \Omega_{ijk} dx^i \wedge dx^j \wedge dx^k$ with $\Omega_{ijk}$ being completely antisymmetric, though of course there is only one independent 3-form in 3D, say $dx^1 \wedge dx^2 \wedge dx^3$ so $\Omega_{ijk}$ is a multiple $\mathcal{J}^{-1}$ of the usual $\eps_{ijk}$ ($\mathcal{J}$ is often called the {\em Jacobian} of the coordinate system).  

The contraction operator $i_B$ is easily written in index notation, e.g.~$i_B\Omega$ is the 2-form with components $B^i\Omega_{ijk}$.  The fact that $\Omega$ is a multiple of $\eps$ is a possible reason for plasma physicists to refer to $\beta$ as the contravariant representation of $B$, but the Jacobian must be taken into account too.

The condition for a volume-form $\Omega$ to be natural for a Riemannian metric $g$ can be written in coordinates as $\Omega = \pm \sqrt{\det g}\ dx^1 \wedge \ldots dx^n$, where $\det g$ is the determinant of the matrix for the covariant representation of $g$ in coordinates $x^1,\ldots x^n$.
The two possible signs for $\Omega$ correspond to the two possible orientations of a connected orientable manifold.

For a {\em simple} $k$-form $f dx^I$, where $I$ is an ordered set of $k$ distinct indices $i_1,\ldots i_k$ and $dx^I = dx^{i_1} \wedge \ldots dx^{i_k}$, the exterior derivative has the formula
\begin{equation}
d(f dx^I) = (\partial_i f) dx^i \wedge dx^I.
\label{eq:d}
\end{equation}
A general $k$-form is a linear combination of simple ones.

On a $k$-form $\omega$,
\begin{equation}
(L_B\omega)_{i_1\ldots i_k} = B^c \partial_c \omega_{i_1\ldots i_k} + (\partial_{i_1}B^c) \omega_{c i_2\ldots i_k} + \ldots + 
(\partial_{i_k} B^c) \omega_{i_1\ldots i_{k-1}c} .
\end{equation}
On a vector field $Y$,
\begin{equation}
L_BY =  (B^j \partial_jY^i - Y^j\partial_j B^i)\partial_i .
\end{equation}

\section{Charged particle motion in a magnetic field}

We now turn to the dynamics of a charged particle in a magnetic field $B$.  Denote its mass by $m$ and its charge by $e$.  The equation of motion in Euclidean space is
\begin{equation}
m \ddot{q} = e \dot{q} \times B(q).
\end{equation}
It is fruitful to put this into Hamiltonian form.  The standard way is to choose a vector potential $A$ for $B$ and introduce a momentum variable $p = m\dot{q}+eA(q,t)$ and let the Hamiltonian be $H(q,p,t) = \frac{1}{2m} |p-eA(q,t)|^2$.  The canonical Hamilton equations $\dot{q}=\partial_p H$, $\dot{p}=-\partial_q H$, reproduce the right equations of motion.

It is better, however, to abandon the canonical view of Hamiltonian systems.  Instead, a Hamiltonian system $\dot{x}=X(x,t)$ (possibly time-dependent) on a manifold $M$ is defined by
\begin{equation}
i_X\omega = dH,
\end{equation}
for a function $H: M \times \R \to \R$ called the {\em Hamiltonian}, and symplectic form $\omega$ on $M$.  A {\em symplectic form} is a non-degenerate closed 2-form.   This equation defines (existence and uniqueness) the vector field $X$ because $\omega$ is non-degenerate.  Note that non-degeneracy of $\omega$ requires $M$ to have even dimension; denoting it by $2n$, we call $n$ the number of {\em degrees of freedom} (DoF). 

The standard example is that $M$ is the cotangent bundle $T^*Q$ of a manifold $Q$, i.e.~the set of covectors to $Q$.  One can write a covector as $(q,p)$ where $q \in Q$ and $p$ is a covector at $q$, i.e.~$p: T_qQ \to \R$ and is linear.
$T^*Q$ has a natural symplectic form, as follows.  
Define the {\em natural 1-form} $\alpha$ on $T^*Q$ by $\alpha_{(q,p)}(\delta q, \delta p) = p(\delta q)$.
In a local coordinate system $q^i$ on $Q$ we define associated coordinates $p_i$ so that $p(\delta q) = p_i {\delta q}^i$ (with summation convention).  Then $\alpha = p_i dq^i$.  
Finally, let $\omega = -d\alpha$.  It is a closed (indeed exact), non-degenerate 2-form.  In the above coordinates $\omega = dq^i \wedge dp_i$.

A {\em simple mechanical system} on $T^*Q$ is defined by this symplectic form $\omega$ and $H(q,p) = \frac12 p^T M^{-1}p + V(q)$ for some positive definite ``mass" matrix $M$, which takes vectors to covectors, and ``potential" $V:Q\to \R$.  Solving $i_X\omega = dH$ for $X=(\dot{q},\dot{p})$ gives
\begin{eqnarray}
\dot{q} &=& M^{-1} p \\
\dot{p} &=& -dV_q ,
\end{eqnarray}
so $M \ddot{q} = -dV_q$.
One can allow $M$ to depend on $q$, which is important to treat mechanical linkages for example, but it adds extra terms to the equations of motion, analogous to centrifugal and Coroilis forces.  An equivalent way to put this is that for a linkage with configuration space $Q$ the kinetic energy is half the norm-squared of the momentum covector for some Riemannian metric $g$ on $Q$, $|p|_q^2 = g^{ij}(q) p_i p_j$ in local coordinates.  

Let us treat the motion of a charged particle in a magnetic field.  We will do it on a general oriented 3D manifold $Q$ with Riemannian metric $g$.  The Riemannian metric gives $|v|^2 = g_q(v,v) = g_{ij}(q)v^i v^j$ for a vector $v$, and for a covector $p$ the length-squared is defined to be $|p|^2 = g^{ij}(q) p_i p_j$, as above.  The dynamics are on the cotangent bundle $T^*Q$.  We take $H=\frac{1}{2m}|p|^2$ and \footnote{If $B$ has a vector potential $A$, or better a 1-form potential $a$, this can be written as $\omega = -d(\alpha + e\pi^*a)$, which does not involve the metric at all.}
\begin{equation}
\omega = -d\alpha - e\pi^*\beta,
\end{equation}
where $\alpha$ is the natural 1-form on $T^*Q$, $\beta = i_B \Omega$, $\pi: T^*Q \to Q$ is $\pi(q,p)=q$ and $\pi^*$ is its pullback from $Q$ to $T^*Q$ (recall (\ref{eq:pullback})).  
So the dynamics $(\dot{q},\dot{p})$ are given by solving 
\begin{equation}
\omega((\dot{q},\dot{p}),(\xi_q,\xi_p)) = dH(\xi_q,\xi_p) \ \forall \xi.  
\end{equation}
Specialising to the case of Euclidean metric, this gives $\dot{q} = p/m$, so $p = m\dot{q}$, the ordinary kinetic momentum, and $-\dot{p}(\xi_q) - e \beta(\dot{q},\xi_q) = 0$ for all $\xi_q$.  Now $\beta(\dot{q},\xi_q) = \Omega(B,\dot{q},\xi_q)$, so this says that $\dot{p}=-eB\times \dot{q}$, which indeed recovers the right equations of motion.

Advantages of the differential forms formulation $i_X\omega = dH$ of Hamiltonian dynamics are that it makes it easy to see that
\begin{enumerate}
\item If $H$ is time-independent then $H$ is conserved along $X$: $i_XdH = i_Xi_X\omega = 0$ by antisymmetry (in the time-dependent case, along solutions we obtain $dH/dt = i_X dH + \partial_t H$, so $dH/dt = \partial_t H$).
\item $\omega$ is conserved along $X$: $L_X\omega = i_Xd\omega + d i_X\omega = 0$ because $d\omega=0$ and $d^2H=0$.
\item It automatically takes care of acceleration in arbitrary coordinate systems (which otherwise requires introducing the Levi-Civita connection into Newton's equations, e.g.~centrifugal and Coriolis forces).
\item It allows to obtain conservation laws and reductions from continuous symmetries.
\end{enumerate}

Let us expand on the latter point.  We say a vector field $U$ on $M$ is a {\em continuous symmetry} of $(H,\omega)$ if $L_UH = 0$ and $L_U \omega = 0$.  Then the second equation shows that $di_U \omega = 0$, thus $U$ is {\em locally Hamiltonian}, i.e.~$i_U\omega = dK$ for some function $K:M\to \R$ locally.  Furthermore, $i_XdK = i_X i_U \omega = -i_U dH = 0$ from the first condition of a symmetry.  So $K$ is conserved by $X$.  This is a version of Noether's theorem.  

There remains the question whether $K$ is globally defined.  Let $\tilde{M}$ be the {\em universal cover} of $M$, i.e.~the set of equivalence classes of curves from a chosen base point, under the equivalence relation of continuous deformation fixing the ends of the curve.  Then the dynamics can be lifted to $\tilde{M}$ and $K$ is globally defined (up to a constant) on $\tilde{M}$.
Next, note that $[X,U]=0$ because
\begin{equation}
i_{[X,U]}\omega = i_X L_U \omega - L_U i_X \omega = 0,
\end{equation}
using $L_U \omega=0$, $i_X\omega=dH$ and $L_UH=0$.  As $\omega$ is nondegenerate, it follows that $[X,U]=0$.  Thus $X$ and $U$ are commuting vector fields on each level set of $K$.
If the first homology group $H_1(M)$ (1D cycles modulo boundaries of 2D surfaces) is spanned by closed trajectories of the set of vector fields of the form $aU + bX$ for arbitrary choices of functions $a,b$ then the change in $K$ around any non-contractible loop is zero and so $K$ is well defined on $M$.  To see this, let $\gamma$ be a closed orbit of $aU+bX$, then 
\begin{equation}
\int_\gamma i_U\omega = \int i_U\omega(\dot{\gamma})\ dt = \int i_U \omega(aU+bX)\ dt = \int \omega(U,aU) + \omega(U,bX)\ dt = 0,
\end{equation}
because of antisymmetry for the first term and the second is $b\ dH(U)=0$.  It follows that the change in $K$ round such a loop is zero.
It is not even necessary to find closed trajectories.  Asymptotic cycles in the sense of Schwartzmann will do \cite{Fri}.

The homology group $H_1(M)$ is dual to the de Rham cohomology group $H^1(M)$ in the sense that given $\gamma \in H_1$ and $\alpha \in H^1$, there is a natural scalar $\langle \alpha,\gamma \rangle = \int_\gamma \alpha$.  It is well-defined despite the freedom to deform $\gamma$ and add any exact 1-form to $\alpha$.  

This version of Noether's theorem is more sophisticated than the usual one for Lagrangian systems, where the symmetry is restricted to being on configuration space.  Hence the need for the additional step of checking whether $K$ is global.

Note the trivial case of $U=X$, which achieves nothing.  If $U,X$ are independent almost everywhere, however, then so are $dK,dH$, so we obtain a genuine reduction of the dynamics by one dimension by restricting to level sets of $K$.

Actually, one can reduce by two dimensions, by also quotienting by the flow $\phi$ of $U$ if its orbit space is a manifold.  The resulting vector field of $X$ on $K^{-1}(k)/\phi$ is Hamiltonian with respect to the reduced symplectic form $\omega$ and Hamiltonian $H$, which we denote by the same symbols.  Note that the symplectic form indeed reduces to the quotient because $\omega(u,\xi)=0$ for all $\xi \in \ker dK$.

Does the symplectic form have physical manifestations?  The answer is yes, e.g.~\cite{Mac}.  One manifestation is that Liouville volume $\Omega = \omega^{\wedge n}/n!$ is conserved by a system of $n$ DoF (here $\omega^{\wedge n}$ is the wedge product of $\omega $ with itself $n$ times, and the factor $n!$ is conventional).  
Since $H$ is also conserved, we deduce conservation of energy-surface volume $\mu_E$ on $H^{-1}(E)$ defined uniquely by taking any $(2n-1)$-form $\mu$ such that $\mu \wedge dH = \Omega$ and restricting it to $H^{-1}(E)$. 
One can write $\mu = i_n\Omega$, with $n = \nabla H/|\nabla H|^2$ for any Riemannian metric.
This is the basis for the theory of entropy for classical mechanical systems.  Another is the action $S = \int_A \omega$ for any area $A$ spanning a closed curve $\gamma$, which is also conserved under the flow of $X$.  Perhaps the action of a closed curve is not considered physical, but bear in mind that if $\gamma$ is a periodic orbit of a Hamiltonian system then it is generically part of a family of such, parametrised by the value $E$ of $H$, and the period $T = dS/dE$.  Furthermore, if the Hamiltonian has slow time-dependence then a periodic orbit drifts in energy in such a way as to preserve its action to high order of approximation (an ``adiabatic invariant").

\section{Charged particle in an axisymmetric magnetic field}

We give an example of application of Noether's theorem to charged particle motion in a magnetic field in Euclidean space.  
Recall that it is Hamiltonian with $H= \frac{1}{2m} |p|^2$ and $\omega = -d\alpha - e\pi^*\beta$ on $T^*\R^3$, where $\alpha$ is the natural 1-form for a cotangent bundle and $\beta = i_B\Omega$.
Let vector field $u=\partial_\phi$ on $T^*\R^3$ with respect to cylindrical coordinates ($r,\phi,z)$ on $\R^3$.  This uses the first-order operator view of a vector field, but we could write it as a velocity field $r\hat{\phi} = (0,r,0, 0,0,0)$ in $(r,\phi,x,p_r,p_\phi,p_z)$. The momentum coordinates are defined so that the natural 1-form $\alpha = p_i dq^i$.  Then $H=\frac{1}{2m} (p_r^2+r^{-2}p_\phi^2+p_z^2)$.  

Say $B$ is {\em axisymmetric} if $L_u\beta=0$ (an alternative definition is $[u,B]=0$, but since $B$ and the chosen $u$ are volume-preserving, this reduces to $L_u\beta=0$).  Then $di_u\beta=0$ because $d\beta=0$.  So $i_u\beta = d\psi$ for some function $\psi$ locally, called a {\em flux function}.  In fact, $\psi$ is global because $\R^3$ is contractible.  But if for some reason the field was defined or axisymmetric only on some axisymmetric solid torus, for example, then $\psi$ would still be global, because 
$i_u d\psi = i_ui_u\beta = 0$ by antisymmetry so $\psi$ is independent of $\phi$.

An alternative approach to deriving a flux function for an axisymmetric magnetic field is to use a vector potential $A$ for $B$ and define axisymmetry by $L_uA^\flat=0$.  Then $i_u\beta = i_udA^\flat = -di_uA^\flat$, so we get a global flux function $\psi = -i_uA^\flat = -A \cdot u = -r A_\phi$.

The relation $i_u\beta = d\psi$ implies $i_ui_B \Omega = d\psi$.  This is written in vector calculus as $B\times u = \nabla \psi$.  Since $u = r\hat{\phi}$ we deduce that
\begin{eqnarray}
B_z &=& -\frac{1}{r}\frac{\partial\psi}{\partial r} \\
B_r &=& \frac{1}{r} \frac{\partial \psi}{\partial z}.
\end{eqnarray}
Note this is a 1DoF Hamiltonian system for the reduction of fieldline flow by axisymmetry: $H=\psi(r,z)$, $\omega = r dz \wedge dr$.
To get the full fieldline flow one just adds $\dot{\phi} = \frac1r B_\phi$.

The axisymmetry of $B$ makes the charged particle dynamics axisymmetric too: $L_u\omega = 0$, $L_u H = 0$.  We determine the resulting constant of the motion by noting that
\begin{equation}
i_u\omega = dp_\phi - ed\psi .
\end{equation}
Thus the motion conserves $L = p_\phi - e\psi$, a modified angular momentum.  
As usual, $p_\phi = mr^2\dot{\phi}$.  So we deduce that $\dot{\phi} = \frac{1}{mr^2}(L+e\psi(r,z))$.
We can reduce the system by axisymmetry to a family of systems on $(r,z,p_r,p_z)$ with $p_\phi -e\psi = L$ constant:
\begin{eqnarray}
H &=& \frac{1}{2m} \left(p_r^2+p_z^2+\frac{(L+e\psi)^2}{r^2}\right) \\
\omega &=& dr \wedge dp_r + dz \wedge dp_z - e B_\phi(r,z) dz \wedge dr.
\end{eqnarray}
This is the basic Hamiltonian system for the motion of charged particles in a tokamak.  To proceed further, however, it is good to notice a further approximate symmetry: gyro-phase rotation.  We will tackle that in the next section, for fields that are not necessarily axisymmetric.

\section{Magnetic moment and guiding-centre motion}

If $B(q(t),t)$ seen by a charged particle changes slowly on the timescale of one gyroperiod $2\pi/\Omega$ ({\em gyrofrequency} $\Omega = -e|B|/m$), then the {\em magnetic moment} $\mu = m|v_\perp|^2/2|B|$ is an adiabatic invariant.  This means that it remains close to its initial value for an exceedingly long time (depending on the smoothness of the variation of $B$).

This is a consequence of the Hamiltonian structure of the dynamics and is most simply revealed using the differential forms approach (as was shown by \cite{Lit} though he used a variational formulation on extended state space).  We treat the time-independent case, but almost no change is required for the time-dependent case.

The idea is to change coordinates from $(q,p)$ to $(X,\rho, p_\pl)$ with {\em guiding centre} $X \in \R^3$, {\em gyroradius vector} $\rho$ perpendicular to $B(X)$ and {\em parallel momentum} $p_\pl \in \R$, given by solving the following system of equations:
\begin{eqnarray}
p &=& e B(X)\times\rho + p_\pl b(X) \\
q &=& X + \rho,
\end{eqnarray}
where $b = B/|B|$.
For $e |B|$ large, there is a unique solution with $\rho$ small (approximately $p\times B/e|B|^2$ evaluated at $q$), by the implicit function theorem.  More precisely, this works if $|\rho|$ is less than the radius of curvature of the fieldlines.

Then the dynamics has approximate rotation symmetry of $\rho$ about $B$.  In particular,
\begin{equation}
H = \frac{|p|^2}{2m} = \frac{1}{2m} (p_\pl^2 + e^2 |B(X)|^2 |\rho|^2)
\end{equation}
is exactly rotation invariant in $\rho$.  The symplectic form requires more work
\begin{equation}
\omega = -d\alpha - e\beta,
\end{equation}
where I have dropped the $\pi^*$ in front of $\beta$ because it is obvious what is meant.
In these $(X,\rho,p_\pl)$ coordinates and choosing perpendicular components $(\rho_1,\rho_2)$ for $\rho$ and denoting the parallel component of $X$ by $X_3$,
\begin{equation}
\alpha = e|B|\rho_2 (dX_1 + d\rho_1) - e |B| \rho_1 (dX_2 + d\rho_2) + p_\pl dX_3.
\end{equation}
The proposed symmetry vector field is $u = (0,0,0,-\rho_2,\rho_1,0)$ in $(X,\rho,p_\pl)$ coordinates.  We will compute $L_u \alpha$ and $L_u \beta$ and then combine them to deduce that $L_u \omega$ is small.

Using Cartan's formula for $L_u$ on differential forms,
\begin{eqnarray}
L_u \rho_2 d\rho_1 &=& i_u d\rho_2 \wedge d\rho_1 + d(-\rho_2^2) = \rho_1 d\rho_1 + \rho_2 d\rho_2 - d(\rho_2^2) \\
L_u (-\rho_1d\rho_2) &=& -i_u d\rho_1 \wedge d\rho_2 - d(\rho_1^2) = \rho_2 d\rho_2 + \rho_1 d\rho_1 - d(\rho_1^2).
\end{eqnarray}
They sum to zero.  Also $L_u (p_\pl dX_3) = 0$ and $L_u |B(X)|=0$.  So we are left with 
\begin{eqnarray}
L_u \alpha &=  & e|B| L_u (\rho_2 dX_1 - \rho_1 dX_2)  \\
&=& e|B|i_u(d\rho_2\wedge dX_1 - d\rho_1 \wedge dX_2) = e|B|(\rho_1dX_1+\rho_2dX_2). \nonumber
\end{eqnarray}
Thus 
\begin{equation}
L_u(-d\alpha) = -e|B(X)|(d\rho_1 \wedge dX_1 + d\rho_2 \wedge dX_2).
\end{equation}

For $L_u \beta$, we have to take into account that $\beta = |B(q)| dq_1 \wedge dq_2$ with $q=X+\rho$.
So 
\begin{equation}
e\beta = e|B(X+\rho)| (dX_1 \wedge dX_2 + dX_1 \wedge d\rho_2 + d\rho_1 \wedge dX_2 + d\rho_1 \wedge d\rho_2).
\end{equation}
Recalling that $\beta$ is closed,
\begin{eqnarray}
&  L_u (-e\beta) = -edi_u \beta = e d\left(|B(q)| (\rho_1dX_1 + \rho_2dX_2 + \rho_2 d\rho_2 + \rho_1 d\rho_1)\right) = \\
&e|B(q)| (d\rho_1 \wedge dX_1  + d\rho_2 \wedge dX_2) + e d|B(q)| \wedge (\rho_1dX_1 + \rho_2dX_2 + \rho_2 d\rho_2 + \rho_1 d\rho_1). \nonumber
\end{eqnarray}
The first term almost cancels $L_u(-d\alpha)$, the difference being just due to where $|B|$ is evaluated, but $\rho$ is small and $B$ varies slowly in space.  The second term is small because it is proportional to $d|B|$ and $B$ varies slowly in space.  Specifically it is
\begin{equation}
e \left( (\rho_2 \partial_{q_1} |B| - \rho_1 \partial_{q_2} |B|)\ dq_1 \wedge dq_2 + \partial_{q_3} |B|\ dX_3 \wedge (\rho_1 dq_1 + \rho_2 dq_2) \right).
\end{equation}
Hence $L_u\omega$ is small and $u$ is an approximate symmetry of the dynamics.

It follows that $i_u\omega$ is approximately $d$ of some function and that function is approximately conserved.  It is conventionally written as $-m\mu/e$ where the {\em magnetic moment}
\begin{equation}
\mu = \frac{e^2}{2m}|B||\rho|^2 = \frac{m|v_\perp|^2}{2|B|}.
\end{equation}
The proof is that, ignoring the dependence of $|B|$ on position, 
\begin{equation}
i_u\omega = -e|B|(\rho_1d\rho_1 + \rho_2d\rho_2) = - \frac{e|B|}{2}d|\rho|^2 = -\frac{m}{e} d\mu.
\end{equation}

Reducing by the approximate symmetry $u$ produces a 2DoF system called {\em first-order guiding-centre motion}, on the space of $(X,p_\pl) \in \R^3 \times \R$:
\begin{eqnarray}
H&=& \frac{1}{2m}p_\pl^2 + \mu |B(X)| \label{eq:Hgc} \\
\omega &=& - d(p_\pl b^\flat) - e \beta_X = b^\flat \wedge dp_\pl - p_\pl db^\flat - e i_B \Omega.
\end{eqnarray}
Let $c = \curl\ b$, so $i_c\Omega = db^\flat$.  Then $\omega$ can be written
\begin{equation}
\omega = b^\flat \wedge dp_\pl - e i_{\widetilde{B}}\Omega,
\end{equation}
with the modified magnetic field
\begin{equation}
\widetilde{B} = B + \frac{p_\pl}{e} c.
\end{equation}
The 2-form $\omega$ is closed; it is non-degenerate except where $\widetilde{B}.b = 0$.
Then the equation of motion is given by solving
\begin{equation}
i_{(\dot{X},\dot{p}_\pl)} \omega = dH.
\end{equation}
Applying to a vector of the form $(0,\delta p_\pl)$, we deduce that
\begin{equation}
\dot{X}_\pl = p_\pl/m.
\label{eq:Xpl}
\end{equation}
Applying to a vector of the form $(\xi,0)$:
\begin{equation}
e (\widetilde{B} \times \dot{X})\cdot\xi = \xi\cdot(\mu \nabla |B| + \dot{p}_\pl b).
\end{equation}
So
\begin{equation}
e\widetilde{B} \times \dot{X} = \mu \nabla |B| + \dot{p}_\pl b.
\label{eq:Xperp}
\end{equation}
Taking the inner product with $\widetilde{B}$ yields
\begin{equation}
\dot{p}_\pl = -\mu \frac{\widetilde{B}\cdot\nabla |B|}{\widetilde{B}\cdot b}.
\label{eq:ppl}
\end{equation}
Finally, taking the cross product of (\ref{eq:Xperp}) with $b$ and using (\ref{eq:Xpl}) we obtain
\begin{equation}
\dot{X} = \frac{1}{\widetilde{B}\cdot b}\left( \frac{\mu}{e}b\times \nabla |B| + \frac{p_\pl}{m} \widetilde{B}\right).
\label{eq:X}
\end{equation}
Equations (\ref{eq:ppl}) and (\ref{eq:X}) are the guiding-centre equations.

One sees there is a repulsion from increasing $|B|$ along $b$ (which corresponds exactly to conservation of $H$ in (\ref{eq:Hgc})), and there is a perpendicular drift  driven by $\mu \nabla_\perp |B|$ and $\frac{p_\pl^2}{me} b\cdot \nabla b$ (coming from the $\curl\ b$ term of $\widetilde{B}$).

There are alternative forms for the guiding centre equations, agreeing to first order in $1/e$, but the advantage of the above approach is that the resulting drift equations retain Hamiltonian structure, allowing deeper understanding of their behaviour and extension to gyrokinetic theory (e.g.~\cite{QCNX}).

\section{Discussion}
We have shown how differential forms can make some results in plasma physics simpler and more intuitive.  They also open up the possibility of new discoveries.

An example is the interaction of two charged particles in a magnetic field.  We found a new constant of the motion in the case where the particles have the same $e/m$ (including sign), corresponding to a novel symmetry that I call ``locomotive coupling rod symmetry'' \cite{MP1,MP2}.  This is an important case because it includes the interaction of two electrons or of two protons.

The topic in plasma physics where I have found differential forms most useful is that of quasi-symmetry.  Quasi-symmetry just means integrability of guiding-centre motion, but without assuming the symmetry corresponds to an isometry.  It is a good principle for stellarator design \cite{Hel}.  Using differential forms, we have gone considerably beyond what was known before \cite{BKM}, though we have not yet a complete understanding of quasi-symmetry.

\section*{Acknowledgements}
I am grateful to Alex Schekochihin for the invitation to write a tutorial on this subject, and to Allen Boozer, Jim Meiss, Nikos Kallinikos and the reviewers for useful comments.  This work was supported by the Simons Foundation /SFARI/(601970, RSM).

\section*{Appendix}

\subsection*{Smooth manifolds}
Here is a slightly more formal introduction to smooth manifolds.  A topological manifold of dimension $n$ is a second countable Hausdorff topological space such that every point has a neighbourhood homeomorphic to $\R^n$ or $\R_+ \times \R^{n-1}$.  A neighbourhood of a point is an open set containing that point.  Second countable means there is a countable set of open sets such that any open set can be obtained by finite intersections and arbitrary unions of them.  Hausdorff means that for any two points $x,y$ there are disjoint neighbourhoods $U_x, U_y$ of $x, y$.  A differentiable manifold $M$ is a topological manifold with an open cover $U_\alpha, \alpha \in A$, and maps $\phi_\alpha: U_\alpha \to \R^n$ which are homeomorphisms to their images such that the change-of-chart maps $\phi_\alpha \circ \phi_\beta^{-1}$ between the subsets of $\R^n$ on which they are defined are diffeomorphisms.  The open cover is called an atlas and the maps $\phi_\alpha$ are called charts.  It is $C^r$ if the change-of-chart maps are $C^r$.

A function $f:M\to \R$ is differentiable if the maps $f \circ \phi_\alpha^{-1}: R^n \to \R$ are differentiable where defined.

\subsection*{Poincar\'e's lemma}
This is the statement that if a $k$-form $\beta$ is closed on a contractible subset $U$ of a manifold then $\beta =d\alpha$ for some $(k-1)$-form on $U$.  Here is a proof.  $U$ contractible means there is a vector field $X$ on $U$ with forward flow $\phi$ that maps $U$ into itself and the image $\phi_tU$ contracts to a point as $t \to \infty$.  Define $(k-1)$-form $\alpha$ on $U$
by
\begin{equation}
\alpha = -\int_0^\infty i_X\phi_t^*\beta\ dt.
\end{equation}
Then 
\begin{equation}
d\alpha = -\int_0^\infty di_X\phi_t^*\beta\ dt = -\int_0^\infty L_X\phi_t^*\beta + i_Xd\phi_t^*\beta,
\end{equation}
using $L_X = di_X + i_X d$ on forms.  The second integrand can be written as $i_X\phi_t^*d\beta$ so is zero by the hypothesis that $\beta$ is closed.  The first integrand can be written as $\partial_t\phi_t^*\beta$, thus $d\alpha$ is minus the change in $\phi_t^*\beta$ from $t=0$ to $\infty$.  But $\phi_t^*\beta \to 0$ as $t\to \infty$ because of the contraction to a point, and $\phi_0^*\beta = \beta$.  Thus $d\alpha = \beta$.

\subsection*{Hodge star}
Although I have chosen not to use it, you may come across the Hodge star operator $\star$ in your reading, so here is a little introduction to it.
In an oriented Riemannian manifold of dimension $n$ it gives a natural bijection between $k$-forms and $(n-k)$-forms.  Let us specialise to $n=3$.  Then any $3$-form $\omega$ is a scalar multiple of the Riemannian volume form $\Omega$, say $\omega = f\Omega$, because the space of top-forms at a point is one-dimensional and $\Omega$ is non-degenerate.  The relation is denoted by $\omega = \star f$, equivalently $f = \star \omega$.  Thus for example one could write $\div\ B = \star d\beta$ (with $\beta = i_B\Omega$).  Similarly, we have seen that a vector field $B$ (not necessarily volume-preserving) produces both a 2-form $\beta = i_B\Omega$ and a 1-form $B^\flat$.  The relation between the two is denoted by $\beta = \star B^\flat$, $B^\flat = \star \beta$.  Thus one can write $J=\curl \ B$ as $J^\flat = \star dB^\flat$ and $\div\ B = \star d \star B^\flat$.  The combination $\delta = \star d \star$ is called the {\em co-differential}, so we could write $\div\ B = \delta B^\flat$.  In particular the Laplacian of a function $f$ is $\Delta f = \delta d f$.  The Laplacian can be extended to $k$-forms by $\Delta = \delta d + d\delta$.

I won't give the general definition of $\star$ here, but it has the nice feature that on odd-dimensional Riemannian manifolds, $\star^2 = 1$, as you have seen in the 3D case.  Beware, however, that in even dimension or for Lorentzian manifolds, $\star^2 = \pm 1$ depending on the degree of the forms involved.

\subsection*{Electromagnetism in space-time}
A beautiful observation is that in a 4D Lorentzian manifold, two of Maxwell's equations can be expressed as the Faraday tensor being a closed 2-form.  Given electric field $E$ and magnetic field $B$, the Faraday tensor is the 2-form
\begin{equation}
F = B_x dy\wedge dz + B_y dz \wedge dx + B_z dx \wedge dy + E_x dx \wedge dt + E_y dy \wedge dt + E_z dz \wedge dt,
\end{equation}
in a local Minkowski coordinate system (i.e.~such that the metric is $ds^2 = dx^2 + dy^2 + dz^2 - c^2 dt^2$ to leading order).
Then Faraday's law ($\frac{\partial B}{\partial t} = -\curl\ E$) and $\div\ B = 0$ are equivalent to the single statement $dF=0$.  This is because 
\begin{equation}
F = \pi^*\beta + (\pi^*E^\flat) \wedge dt, 
\end{equation}
where $\pi$ maps space-time to space, so using (\ref{eq:d}), $dF=0$ iff $d\beta=0$ and $\frac{\partial\beta}{\partial t} + dE^\flat = 0$.

The electromagnetic force on a charge $e$ with 4-velocity $U$ in space-time is the 1-form $f = -e i_U F$.  One can do a relativistic treatment of the reduction of charged particle motion to guiding-centre motion.

Using the Hodge star, the two remaining Maxwell equations can be expressed as
\begin{equation}
d\star F = \star J,
\label{eq:dstarF}
\end{equation}
where $J$ is a 1-form on space-time representing densities of charge and current.  Applying $d$ to (\ref{eq:dstarF}) yields $d\star J=0$, which expresses charge conservation.
The beauty of the formulation is that it applies to arbitrary Lorentzian manifolds, taking care automatically of the effects of curvature.

\subsection*{Helicity}
The helicity integral $\mathcal{H} = \int A\cdot B\ dV$, where $A$ is a vector potential for $B$, plays an important role in Taylor's theory of relaxation of plasma in a magnetic field and also can be interpreted as an average rate of winding of field lines around each other \cite{AK}.  It has a nice representation in differential forms, namely
\begin{equation}
\mathcal{H} = \int a \wedge da ,
\end{equation}
where $a = A^\flat$ is the 1-form potential for $B$.
This integrand is a baby version of the famous Chern-Simons form on 3-manifolds, where the concept of differential form is extended to take values in a Lie algebra instead of just $\R$ \cite{AK}.

\subsection*{Lie derivative on general tensors}
A tensor $T$ of type $(p,q)$ is a multilinear map from the space of $p$ covectors and $q$ vectors at a point to $\R$.
Its Lie derivative along a vector field $Y$ can be computed by
\begin{align}
&(L_YT) (\alpha_1,\ldots,\alpha_p,X_1,\ldots X_q) = Y(T (\alpha_1,\ldots,\alpha_p,X_1,\ldots X_q)) \\
&- T(L_Y\alpha_1,\alpha_2,\ldots X_q) - T(\alpha_1,L_Y\alpha_2,\ldots, X_q) - \ldots \nonumber \\
&- T(\alpha_1,\ldots \alpha_p,L_Y X_1,X_2,\ldots X_q) - \dots - T(\alpha_1,\ldots X_{q-1}, L_YX_q), \nonumber
\end{align}
for any 1-forms $\alpha_i$ and vector fields $X_j$.


\begin{thebibliography}{WWW}
\bibitem[Arn]{Arn} Arnol'd VI, Mathematical methods of classical mechanics (Springer, 1978)
\bibitem[AK]{AK} Arnol'd VI, Khesin BA, Topological methods in hydrodynamics (Springer, 1998)
\bibitem[BKM]{BKM} Burby JW, Kallinikos N, MacKay RS, Some mathematics for quasi-symmetry, arXiv:1912.06468 
\bibitem[CL]{CL} Cary JR, Littlejohn RG, Noncanonical Hamiltonian mechanics and its application to magnetic field line flow, Ann Phys 151 (1983) 1--34.
\bibitem[De]{De} Deschamps GA, Electromagnetics and differential forms, Proc IEEE 69:6 (1981) 676--87.
\bibitem[Fri]{Fri} Fried D, The geometry of cross sections to flows, Topology 21 (1982) 353--71.
\bibitem[Hel]{Hel} Helander P, Theory of plasma confinement in non-axisymmetric magnetic fields, Rep Prog Phys 77 (2014) 087001.
\bibitem[Lit]{Lit} Littlejohn RG, Variational principles of guiding centre motion, J Plasma Phys 29 (1983) 111--25.
\bibitem[Mac]{Mac} MacKay RS, Some Aspects of the Dynamics and Numerics of Hamiltonian Systems, in: The Dynamics of Numerics and Numerics of Dynamics, eds Broomhead DS, Iserles A, IMA conf proc (Oxford, 1992) 137--93.
\bibitem[MP1]{MP1} MacKay RS, Pinheiro D, Interaction of two charges in a uniform magnetic field: I. Planar problem, Nonlinearity 19 (2006) 1713--45.
\bibitem[MP2]{MP2} MacKay RS, Pinheiro D, Interaction of two charges in a uniform magnetic field: II spatial case, J Nonlin Sci 18 (2008) 615--66.
\bibitem[MTW]{MTW} Misner CW, Thorne KS, Wheeler JA, Gravitation (Freeman, 1973)
\bibitem[QCNX]{QCNX} Qin H, Cohen RH, Nevins WM, Xu XQ, Geometric gyrokinetic theory for edge plasmas, Phys Plasmas 14 (2007) 056110.
\bibitem[WR]{WR} Warnick KF, Russer PH, Differential forms and electromagnetic field theory, Prog Electromagnetic Res 148 (2014) 83--112.
\end{thebibliography}
\end{document}